\documentclass[runningheads,a4paper]{llncs}

\usepackage{cite}
\usepackage[utf8]{inputenc}
\usepackage{graphicx}
\usepackage{amsmath}
\usepackage{hyperref}
\usepackage{amsfonts}
\usepackage{array}
\usepackage{caption}
\usepackage{amssymb}
\usepackage[english]{babel}
\usepackage{appendix}
\usepackage[linesnumbered,ruled]{algorithm2e}
\newcommand*{\QEDB}{\hfill\ensuremath{\square}}

\usepackage{amsthm}
\theoremstyle{plain}
 \usepackage{tikz}
\usetikzlibrary{arrows,patterns,positioning}
\newcommand*\circled[1]{\tikz[baseline=(char.base)]{
            \node[shape=circle,draw,inner sep=1pt] (char) {#1};}}
\usepackage{url}
\urldef{\mailsa}\path|{muhammad.syifaul.mufid, alessandro.abate}@cs.ox.ac.uk|    
\urldef{\mailsb}\path| dieky@matematika.its.ac.id|
\newcommand{\keywords}[1]{\par\addvspace\baselineskip
\noindent\keywordname\enspace\ignorespaces#1}

\begin{document}

\mainmatter  

\title{Tropical Abstractions of Max-Plus Linear Systems}

\titlerunning{Tropical Abstractions of Max-Plus Linear Systems}

%
%
\author{Muhammad Syifa'ul Mufid$^\dagger$%
\and Dieky Adzkiya$^\ddagger$
\and Alessandro Abate$^\dagger$}
\authorrunning{Muhammad Syifa'ul Mufid, Dieky Adzkiya, Alessandro Abate}

\institute{$^\dagger$Department of Computer Science, University of Oxford\\United Kingdom\\
\mailsa\vspace*{3ex}\\
$^\ddagger$Department of Mathematics, Institut Teknologi Sepuluh Nopember\\
 Surabaya, Indonesia\\
\mailsb\\
}

%
%

\maketitle

\begin{abstract}
This paper describes the development of finite abstractions of Max-Plus-Linear (MPL) systems using tropical operations. 
The idea of tropical abstraction is inspired by the fact that an MPL system is a discrete-event model updating its state with operations in the tropical algebra. 
The abstract model is a finite-state transition system: we show that the abstract states can be generated by operations on the tropical algebra, 
and that the generation of transitions can be established by tropical multiplications of matrices. 
The complexity of the algorithms based on tropical algebra is discussed and their performance is tested on a numerical benchmark against an existing alternative abstraction approach. 
\keywords{MPL system, tropical algebra, definite form, difference-bound matrix, abstraction, reachability}
\end{abstract}

\section{Introduction}
Tropical mathematics is a rapidly growing subject since it was firstly introduced \cite{Pin}. It has branches in mathematical fields such as tropical geometry \cite{Grigory} and tropical algebra \cite{Pin}. The latter denotes an algebraic structure that uses max or min for addition and + for multiplication, respectively - hence, it is well known as max-plus or min-plus algebra. In this paper, we use the former operation to define the tropical algebra. 

A class of discrete-event system (DES) based on tropical algebra is the Max-Plus-Linear (MPL) one \cite{Bacelli}. 
Models of MPL systems involve tropical operations, namely max and +. The state space of these models represents the timing of events that are synchronised over the max-plus algebra. 
This means that the next event will occur right after the last of the previous events has finished. 
The application of MPL systems is significantly found on models where time variable is essential such as transportation networks \cite{Hiedergott}, scheduling \cite{Alirezaei}, and manufacturing \cite{Aleksey}. Another MPL application deals with biological systems \cite{Chris}. 


Formal abstractions denote a set of techniques to generate abstract versions of large or even infinite models \cite{Baier}. 
This results in less complex {abstract models}, which allow to replace the analysis of the original or concrete ones with automated and scalable techniques. 
The abstract states and abstract transitions are generated based on a so-called {abstraction function}. 
Often the relation between concrete and abstract model can be formalised by the notion of simulation \cite{Baier}. 

Finite abstractions of MPL system have been firstly introduced in \cite{Dieky1}. 
These abstraction procedures start by transforming a given MPL system into a Piece-Wise Affine (PWA) model \cite{Heemels}. 
The PWA model is characterised by several domains (partitions, or PWA regions) and corresponding affine dynamics.  
The resulting abstract states are the partitions corresponding to the PWA regions. 
Finally, the transition relation between pairs of abstract states depends on the trajectory of the original MPL system. 
This abstraction technique enables one to perform model checking over an MPL system; one of the applications is safety analysis \cite{Dieky1}. 
Interested readers are referred to \cite{Dieky1,Dieky2,Dieky3} and the \textsf{VeriSiMPL} toolbox \cite{Dieky4}.

This paper introduces the idea of Tropical Abstractions of MPL systems. 
The approach is inspired by the fact that an MPL system is a DES that is natively updated via tropical operations.  
We will show that the abstraction of MPL systems can be established by tropical operations and with algorithms exclusively based on tropical algebra. 
We argue by experiments that this has clear computational benefits on existing abstraction techniques. 

The paper is outlined as follows. Section 2 is divided into three parts. 
The first part explains the basic of MPL systems including the properties of its state matrix. 
We introduce the notion of region matrix and of its conjugate, which play a significant role in the abstraction procedures.  The notion of definite form and its generalisation are explained in the second part.  Finally, we introduce a new definition of Difference Bound Matrices (DBM) \cite{Dill}. 

Equipped with these notions, all algorithms of the tropical abstraction procedure are explained in Section 3. In particular, we prove that the the resulting PWA regions characterised by the MPL system are equivalent to the definite form of the state matrix. 
We also show that both computation of image and inverse image can be established with tropical matrix multiplications w.r.t. the region matrix and its conjugate -- this is later used for reachability analysis (forward and backward).  The comparison of the algorithms performance against the state of the art is presented in Section 4. The paper is concluded with Section 5. 
The proofs of the results are in the Appendix. 


\section{Models and Preliminaries}

This section discusses the notion of Max-Plus Linear systems \cite{Bacelli} and the definite form of tropical matrices \cite{Sergey}, then it introduces the concept of Difference-Bound Matrices (DBM) as tropical matrices.

\subsection{Max-Plus-Linear Systems}

In tropical algebra, $\mathbb{R}_{\max}$ is defined as $\mathbb{R}\cup \{-\infty\}$. 
This set is equipped with two binary operations, $\oplus$ and $\otimes$, where 
$$
a\oplus b := \max\{a,b\}~~\textrm{and}~~a\otimes b:=a+b,  
$$
for all $a,b\in \mathbb{R}_{\max}$. 
The algebraic structure $(\mathbb{R}_{\max},\oplus,\otimes) $ is a semiring with $\varepsilon:=-\infty$ and $e:=0$ as the null and unit element, respectively \cite{Bacelli}.

The notation $\mathbb{R}_{\max}^{m\times n}$ represents the set of $m\times n$ tropical matrices whose elements are in $\mathbb{R}_{\max}$. 
Tropical operations can be extended to matrices as follows. If $A,B\in  \mathbb{R}_{\max}^{m\times n}, C\in \mathbb{R}_{\max}^{n\times p}$ then
\begin{eqnarray}
\nonumber [A\oplus B](i,j) &=& A(i,j)\oplus B(i,j)\\
\nonumber [A\otimes C](i,j) &=& \bigoplus_{k=1}^n A(i,k)\otimes C(k,j)
\end{eqnarray}
for all $i,j$ in the corresponding dimension. Given a natural number $m$, the tropical power of $A\in \mathbb{R}_{\max}^{n\times n}$ is denoted by $A^{\otimes m}$ and corresponds to $A\otimes \ldots \otimes A$ ($m$ times). As we find in standard algebra, the zero power $A^{\otimes 0}$ is an $n\times n$ identity matrix $I_n$, where all diagonals and non-diagonals are $e$ and $\varepsilon$, respectively. 
%

An (autonomous) MPL system is defined as
\begin{equation}
x(k+1) = A\otimes x(k),
\end{equation}
where $A\in \mathbb{R}_{\max}^{n\times n}$ is the matrix system and $x(k) = [x_1(k)\ldots x_n(k)]^\top$ is the state variables \cite{Bacelli}. 
Traditionally, $x$ represents the time stamps of the discrete-events, while $k$ corresponds to an event counter. 

\begin{definition}[Precedence Graph \cite{Bacelli}]
The precedence graph of $A$, denoted by $\mathcal{G}(A)$, is a weighted directed graph with nodes $1,\ldots,n$ and an edge from $j$ to $i$ with weight $A(i,j)$ if $A(i,j)\neq \varepsilon$. The weight of a path $p=i_1i_2\ldots i_k$ is equal to the total weight of the corresponding edges i.e. $w(p)=A(i_2,i_1)+\ldots+A(i_{k},i_{k-1})$.
\end{definition}
\begin{definition}[Regular (Row-Finite) Matrix \cite{Hiedergott}] A matrix $A\in \mathbb{R}_{\max}^{n\times n}$ is called regular (or row-finite) if there is at least one finite element in each row.
\end{definition}
The following notations deal with a row-finite matrix $A\in\mathbb{R}_{\max}^{n\times n}$. The coefficient $g=(g_1,\ldots,g_n)\in\{1,\ldots,n\}^n$ is called \textit{finite coefficient} iff $A(i,g_i)\neq \varepsilon$ for all $1\leq i\leq n$. We define the \textit{region matrix} of $A$ w.r.t. the finite coefficient $g$ as
\begin{equation}
A_g(i,j)=\left\{
\begin{array}{cl}
A(i,j), &\textrm{if}~g_i=j\\
\varepsilon,&\textrm{otherwise}.
\end{array}
\right.
\label{eq0}
\end{equation}
One can say that $A_g$ is a matrix that keeps the finite elements of $A$ indexed by $g$. The \textit{conjugate} of $A$ is $A^\mathsf{c}$, where
\begin{equation}
A^\mathsf{c}(i,j) = \left\{
\begin{array}{cl}
-A(j,i),&\text{if}~A(i,j) \neq \varepsilon\\
\varepsilon,&\text{otherwise}.\\
\end{array}\right.
\label{eq2}
\end{equation}

\subsection{Definite Forms of Tropical Matrices}

The {concept} of \textit{definite form} over a tropical matrix was firstly introduced in \cite{Sergey}. 
Consider a given $A\in \mathbb{R}_{\max}^{n\times n}$ and let $\alpha$ be one of the maximal permutations\footnote{A permutation $\alpha$ is called maximal if $\bigotimes_{i=1}^n A(i,\alpha(i))=\text{per}(A)$, where $\text{per}(A)$ is the permanent of $A$ \cite{Peter,Sergey}.} of $A$. The definite form of $A$ w.r.t. $\alpha$ is $\overline{A}_\alpha$, where
\begin{equation}
\overline{A}_\alpha(i,j) = A(i,\alpha(j))\otimes A(j,\alpha(j))^{\otimes -1} = A(i,\alpha(j))- A(j, \alpha(j)).
\label{eq4}
\end{equation}

In this paper, we allow for a generalisation of the notion of definite form. We generate the definite form from the finite coefficients introduced above. Notice that the maximal permutation is a special case of finite coefficient $g=(g_1,\ldots,g_n)$ when all $g_i$ are different. 
Intuitively, the definite form over a finite coefficient $g$ is established by; 1) column arrangement of $A$ using $g$ i.e. $B(\cdot,j)=A(\cdot,g_j)$ and then 2) subtracting each column by the corresponding diagonal element i.e. $\overline{A}_g(\cdot,j)= B(\cdot,j)-B(j,j)$ for all $j\in\{1,\ldots,n\}$. 

Furthermore, we define two types of definite forms. We call the definite form introduced in \cite{Sergey} to be a \textit{column-definite} form. We define as an additional form the \textit{row-definite} form $_g\overline{A}$. The latter form is similar to the former, except that now the row arrangement is used, 
namely $B(g_i,\cdot)=A(i,\cdot)$ for all $i\in\{1,\ldots,n\}$. 
Notice that, in a row arrangement, one could find two or more different rows of $A$ are moved into the same row at $B$. 
As a consequence, some rows of $B$ remain empty. 
In these cases, $\varepsilon$ is used to fill the empty rows. 
For rows with multiple entries, we take the maximum point-wise after subtracting by the corresponding diagonal element. 

%
\begin{example}
Consider a tropical matrix $$A=\begin{bmatrix}
~\varepsilon&~1&~3\\
~5&~\varepsilon&~4\\
~7&~8&~\varepsilon
\end{bmatrix}.$$
and a finite coefficient $g=(2,1,1)$. The row-definite form for $g$ is
$$
A=\begin{bmatrix}
\varepsilon&~\circled{1}&~3\\
\circled{5}&~\varepsilon&~4\\
\circled{7}&~8&~\varepsilon
\end{bmatrix}\dashrightarrow\left[
\begin{array}{ccc}
{5}&~\varepsilon&~4\\
{7}&~8&~\varepsilon\\\hline
\varepsilon&~{1}&~3\\\hline
\varepsilon&~\varepsilon&~\varepsilon
\end{array}
\right]\dashrightarrow \left[\begin{array}{ccc}
0&~\varepsilon&~-1\\
0&~1&~\varepsilon\\\hline
\varepsilon&~0&~2\\\hline
\varepsilon&~\varepsilon&~\varepsilon
\end{array}
\right]\dashrightarrow ~_g\overline{A}=\left[\begin{array}{ccc}
0&~1&~-1\\
\varepsilon&~0&~2\\
\varepsilon&~\varepsilon&~\varepsilon
\end{array}
\right].
$$
On the other hand, the column-definite form w.r.t. $g$ is
$$
A=\begin{bmatrix}
\varepsilon&~\circled{1}&~3\\
\circled{5}&~\varepsilon&~4\\
\circled{7}&~8&~\varepsilon
\end{bmatrix}\dashrightarrow\left[
\begin{array}{cc|cc|c}
1&&~\varepsilon&&~\varepsilon\\
\varepsilon&&~5&&~5\\
8&&~7&&~7
\end{array}
\right]\dashrightarrow \overline{A}_g=\left[\begin{array}{ccc}
0&~\varepsilon&~\varepsilon\\
\varepsilon&~0&~-2\\
7&~2&~0
\end{array}
\right].
$$
\end{example}
\begin{proposition}
\label{prop2}
The column-definite and row-definite form of $A\in \mathbb{R}_{\max}^{n\times n}$ w.r.t. a finite coefficient $g$ are  $\overline{A}_g=A\otimes A_g^\mathsf{c}$ and $_g\overline{A}=A_g^\mathsf{c}\otimes A $, respectively.
\end{proposition}

\subsection{Difference Bound Matrices as Tropical Matrices}

This section discusses the idea of treating Difference Bound Matrices as tropical matrices, and some related properties.

\begin{definition}[Difference Bound Matrices]
A DBM in $\mathbb{R}^n$ is the intersection of sets defined by $x_i-x_j\sim_{i,j} d_{i,j}$, 
where $\sim_{i,j}\in \{>,\geq\}$ and $d_{i,j}\in \mathbb{R}\cup\{-\infty\}$ for $0\leq i,j\leq n$. 
The variable $x_0$ is set to be equal to 0. 
\label{def4}
\end{definition}

\noindent The dummy variable $x_0$ is used to allow for the single-variable relation $x_i\sim c$, which can be written as $x_i-x_0\sim c$. \autoref{def4} slightly differs from \cite{Dill} as we use operators $\{>,\geq\}$ instead of $\{<,\leq\} $. The reason for this alteration is to transfer DBMs into the tropical domain. 

A DBM in $\mathbb{R}^n$ can be expressed as a pair of matrices $(D,S)$. The element $D(i,j)$ stores the bound variable $d_{i,j}$, while $S(i,j)$ represents the \textit{sign matrix} of the operator i.e. $S(i,j)=1$ if $\sim_{i,j}~=~\geq $ and $S(i,j)=0$ otherwise. In the case of $i=j$, it is more convenient to put $D(i,i)=0$ and $S(i,i)=1$, as it corresponds to $x_i-x_i\geq 0$. 

Notice that, under \autoref{def4}, each DBM $D$ in $\mathbb{R}^n$ is an $(n+1)$-dimensional tropical matrix. Throughout this paper, we may not include the sign matrix whenever recalling a DBM.
Some operations and properties in tropical algebra can be used for DBM operations, such as intersection, computation of the canonical form, and emptiness checking. 
Such DBM operations are key for developing abstraction procedures. 

\begin{proposition}
The intersection of DBM $D_1$ and $D_2$ is equal to $D_1\oplus D_2$.\QEDB
\end{proposition}
The sign matrix for $D_1\oplus D_2$ is determined separately as it depends on the operator of the tighter bound. 
More precisely, suppose that $S_1,S_2$ and $S$ are the sign matrices of $D_1,D_2$ and of $D_1\oplus D_2$ respectively, then
$$
S(i,j)=\left\{
\begin{array}{ll}
S_1(i,j),&\text{if}~D_1(i,j)>D_2(i,j)\\
S_2(i,j),&\text{if}~D_1(i,j)<D_2(i,j)\\
\min\{S_1(i,j),S_2(i,j)\},&\text{if}~D_1(i,j)=D_2(i,j).
\end{array}
\right.
$$
\indent  
Any DBM admits a graphical representation, called the potential graph, interpreting the DBM $D$ as a weighted directed graph \cite{Peron}. Because each DBM is also a tropical matrix, the potential graph of $D$ can be viewed as a precedence graph $\mathcal{G}(D)$.

The canonical-form of a DBM $D$, denoted as $\mathsf{cf}(D)$, is a DBM with the tightest possible bounds \cite{Dill}. The advantage of the canonical-form representation is that emptiness checking can be evaluated very efficiently. Indeed, for a canonical DBM $(D,S)$, if there exist $0\leq i\leq n$ such that $D(i,i)>0$ or $S(i,i)=0$ then the DBM corresponds to an empty set.  Computing the canonical-form representation is done by the all-pairs shortest path (APSP) problem over the corresponding potential graph \cite{Dill,Peron}. (As we alter the definition of the DBM, it is now equal to all-pairs longest path (APLP) problem.) One of the prominent algorithms is Floyd-Warshall \cite{Floyd} which has a cubic complexity w.r.t. its dimension. 


On the other hand, in a tropical algebra sense, $[D^{\otimes m}](i,j)$ corresponds to the maximal total weights of a path with length $m$ from $j$ to $i$ in $\mathcal{G}(D)$. Furthermore, $[\bigoplus_{m=0}^{n+1} D^{\otimes m}](i,j)$ is equal to the maximal total weights of a path from $j$ to $i$. Thus, $\bigoplus_{m=0}^{n+1} D^{\otimes m} $ is indeed the solution of APLP problem. 
\autoref{prop4a} provides an alternative computation of the canonical form of a DBM $D$ based on tropical algebra. 
\autoref{prop4b} relates non-empty canonical DBMs with the notion of definite matrix. A tropical matrix $A$ is called definite if $\text{per}(A)=0$ and all diagonal elements of $A$ are zero \cite{Peter}.
\begin{proposition}
\label{prop4a}
Given a DBM $D$, the canonical form of $D$ is $\mathsf{cf}(D)=\bigoplus_{m=0}^{n+1} D^{\otimes m}$, 
where $n$ is the number of variables excluding $x_0$. \QEDB
\end{proposition}
\begin{proposition}
\label{prop4b}
Suppose $D$ is a canonical DBM. If $D$ is not empty then it is definite.\QEDB
\end{proposition}
\section{MPL Abstractions Using Tropical Operations}
This section introduces the concept of \textit{tropical abstractions}. 
Firstly, the comparison with the abstraction method in \cite{Dieky1} is described. 
Then, we provide a new procedure to generate abstract states and transitions based on tropical algebra. 
\subsection{Related Work}
The notion of abstraction of an MPL system has been first discussed in \cite{Dieky1}. 
The procedure starts by transforming the MPL system characterised by $A\in \mathbb{R}_{\max}^{n\times n}$ into a PWA (piece-wise affine) model \cite[Algorithm 2]{Dieky1}, 
and then considering the partitions associated to the obtained PWA \cite[Algorithm 6]{Dieky1}. 
The abstract states associated to the partitions are represented by DBMs. 
The transitions are then generated using one-step forward-reachability analysis \cite{Dieky1}: 
first, the image of each abstract state w.r.t. the MPL system is computed; 
then, each image is intersected with partitions associated to other abstract states; 
finally, transition relations are defined for each non-empty intersection. 
This procedure is summarised in \cite[Algorithm 7]{Dieky1}. 

The computation of image and of inverse image of a DBM is described in \cite{Dieky3}. 
These computations are used to perform forward and backward reachability analysis, respectively. 
The worst-case complexity of both procedures is $O(n^3)$, where $n$ is the number of variables in $D$ excluding $x_0$. 
A more detailed explanation about image and inverse image computation of a DBM is in Section 3.3. 

\subsection{Generating the Abstract States}

We begin by recalling the PWA representation of an MPL system characterised by a row-finite matrix $A\in \mathbb{R}_{\max}^{n\times n}$. 
It is shown in \cite{Heemels} that each MPL system can be expressed as a PWA system. 
The PWA system comprises of convex domains (or PWA regions) and has correspondingly affine dynamics. 
The PWA regions are generated from the coefficient $g=(g_1,\ldots,g_n)\in \{1,\ldots,n\}^n$. As shown in \cite{Dieky1}, the PWA region corresponding to coefficient $g$ is
\begin{equation}
R_g = \bigcap_{i=1}^n\bigcap_{j=1}^n\left\{\textbf{\text{x}}\in \mathbb{R}^n|x_{g_i}-x_j\geq A(i,j)-A(i,g_i)\right\}.
\label{eq7}
\end{equation}
Notice that, if $g$ is not a finite coefficient, then $R_g$ is empty. 
However, a finite coefficient might lead to an empty set. Recall that the DBM $R_g$ in \eqref{eq7} is not always in canonical form.
\begin{definition}[Adjacent Regions {\cite[Def. 3.10]{Dieky1}}] 
Suppose $R_{g}$ and $R_{g^\prime}$ are non-empty regions generated by \eqref{eq7}. These regions are called adjacent, denoted by $R_{g}>R_{g^\prime}$, if there exists a single $i\in\{1,\ldots,n\}$ such that $g_i>g^\prime_i$ and $g_j= g^\prime_j$ for each $j\neq i$.
\end{definition}
The affine dynamic of a non-empty $R_g$ is
\begin{equation}
x_i(k+1)=x_{g_i}(k)+A(i,g_i),~~ i=1,\ldots,n.
\label{eq8}
\end{equation}
Notice that Equation \eqref{eq8} can be expressed as $x(k+1)=A_g\otimes x(k)$, where $A_g$ is a region matrix that corresponds to a finite coefficient $g$. As mentioned before, a PWA region $R_g$ is also a DBM. The DBM $R_g$ has no dummy variable $x_0$. For simplicity, we are allowed to consider $R_g$ as a matrix, that is $R_g\in \mathbb{R}_{\max}^{n\times n}$. We show that $R_g$ is related to the row-definite form w.r.t. the finite coefficient $g$.

\begin{proposition}
For each finite coefficient $g$, $R_g=~_g\overline{A}\oplus I_n$. \QEDB
\end{proposition}
Algorithm 1 provides a procedure to generate the PWA system from a row-finite $A\in\mathbb{R}_{\max}^{n\times n}$. It consists of: 1) generating region matrices (line 3) and their conjugates (line 4), 2) computing the row-definite form (line 5), and 3) emptiness checking of DBM $R_g$ (lines 6-7). The first two steps are based on tropical operations while the last one is using the Floyd-Warshall algorithm. The complexity of Algorithm 1 depends on line 6; that is $O(n^3)$. The worst-case complexity of Algorithm 1 is $O(n^{n+3})$ because there are $n^n$ possibilities at line 1. However, we do not expect to incur this worst-case complexity, especially when a row-finite $A$ has several $\varepsilon$ elements in each row.

In \cite{Dieky1}, the abstract states are generated via refinement of PWA regions. Notice that, for each pair of adjacent regions $R_g$ and $R_{g^\prime}$, $R_g\cap R_{g^\prime}\neq \emptyset$. The intersection of adjacent regions is removed from the region with the lower index. Mathematically, if $R_g>R_{g^\prime}$ then $R_{g^\prime}:=R_{g^\prime} \setminus R_g$. 

Instead of removing the intersection of adjacent regions, the partition of PWA regions can be established by choosing the sign matrix for $R_g$ i.e. $S_g$. As we can see in \eqref{eq7}, all operators are $\geq$. Thus, by \eqref{eq7}, $S_g(i,j)=1$ for all $i,j\in\{1,\ldots,n\}$. In this paper, we use a rule to decide the sign matrix of $R_g$ as follows
\begin{equation}
S_g(i,j)=\left\{
\begin{array}{ll}
1,&\text{if}~R_g(i,j)> 0~\text{or}\\
&~~~R_g(i,j)= 0~\text{and}~i\leq j,\\
0,&\text{if}~R_g(i,j)< 0~\text{or}\\
&~~~R_g(i,j)= 0~\text{and}~i>j.
\end{array}\right.
\label{eq9}
\end{equation}
This rule guarantees empty intersection for each pair of region.
\newpage
\begin{algorithm}[t]
\label{alg1}
    \SetKwInOut{Input}{Input}
    \SetKwInOut{Output}{Output}
    \Input{$A\in\mathbb{R}_{\max}^{n\times n}$, a row-finite tropical matrix}
    \Output{\textbf{R},\textbf{A}, a PWA system over $\mathbb{R}^n$\\
    where \textbf{R} is a set of regions and \textbf{A} represent a set of affine dynamics
    }
      \For{$g\in\{1,\ldots,n\}^n$ }{
		\If{$g~\mathrm{is~a~finite~coefficient}$}
      {
        generate $A_g$ according to \eqref{eq0} \\
        generate $A_g^\mathsf{c}$ from $A_g$ according to \eqref{eq2}\\
		$R_g:=(A_g^\mathsf{c}\otimes A)\oplus I_n$\\
		$R_g:=\mathsf{cf}(R_g)$\\
		\If{$R_g~\mathrm{is~not~empty}$}
		{
		$\textbf{R}:=\textbf{R}\cup \{R_g\},\textbf{A}:=\textbf{A}\cup \{A_g\}$
		}
      }
      
    }
    \caption{Generating the PWA system using tropical operations}
\end{algorithm}
\vspace*{-6ex}
Algorithm 2 is a modification of Algorithm 1 by applying rule in \eqref{eq9} before checking the emptiness of $R_g$. Notation $R_g:=(R_g,S_g)$ in line 7 is to emphasise that DBM $R_g$ is now associated with $S_g$. It generates the partitions of PWA regions which represent the abstract states of an MPL system characterised by $A\in\mathbb{R}_{\max}^{n\times n}$. The worst-case complexity of Algorithm 2 is similar to that of Algorithm 1.
\vspace*{-4ex}
\begin{algorithm}[!ht]
    \caption{Generating a partition from region of PWA system by tropical operations}
\label{alg2}

    \SetKwInOut{Input}{Input}
    \SetKwInOut{Output}{Output}
    \Input{$A\in\mathbb{R}_{\max}^{n\times n}$, a row-finite tropical matrix}
    \Output{\textbf{R},\textbf{A}, a PWA system over $\mathbb{R}^n$\\
    where \textbf{R} is a set of regions and \textbf{A} represent a set of affine dynamics
    }
      \For{$g\in\{1,\ldots,n\}^n$ }{
		\If{$g~\mathrm{is~a~finite~coefficient}$}
      {
        generate $A_g$ according to \eqref{eq0} \\
        generate $A_g^\mathsf{c}$ from $A_g$ according to \eqref{eq2}\\
		$R_g:=A_g^\mathsf{c}\otimes A$\\
		generate sign matrix $S_g$ from $R_g$ according to \eqref{eq9}\\
		$R_g:=(R_g,S_g)$\\
		$R_g:=\mathsf{cf}(R_g)$\\
		\If{$R_g~\mathrm{is~not~empty}$}
		{
		$\textbf{R}:=\textbf{R}\cup \{R_g\},\textbf{A}:=\textbf{A}\cup \{A_g\}$
		}
      }
      
    }
\end{algorithm}
\vspace*{-6ex}
\begin{remark} 
The resulted $R_g$ in Algorithm 1 and Algorithm 2 is an $n$-dimensional matrix which represents a DBM without dummy variable $x_0$. This condition violates Definition \ref{def4}. To resolve this, the system matrix $A\in\mathbb{R}_{\max}^{n\times n}$ is extended into $(n+1)$-dimensional matrix by adding the $0^\text{th}$ row and column as follows
\[A(0,\cdot)=[0 ~~\varepsilon ~~\ldots ~~\varepsilon],~~A(\cdot,0)=A(0,\cdot)^\top.\] 
As a consequence, the finite coefficient $g$ is now an $(n+1)$-row vector $g=(g_0,g_1,\ldots,g_n)$ where $g_0$ is always equal to 0. For the rest of this paper, all matrices are indexed starting from zero. \QEDB
\end{remark}

As explained in \cite{Dieky1}, each partition of PWA regions is treated as an abstract state. Therefore, the number of abstract states is equivalent to the cardinality of partitions. Suppose $\hat{R}$ is the set of abstract states, then $\hat{R}$ is a collection of all non-empty $R_g$ generated by Algorithm 2. 
\subsection{Image and Inverse Image Computation of DBMs}
This section describes a procedure to compute the image of DBMs w.r.t. affine dynamics. First, we recall the procedures from \cite{Dieky3}. Then, we develop new procedures based on tropical operations. The proofs of the results are in the Appendix.

The image of a DBM $D$ is computed by constructing a DBM $\textbf{D}$ consisting of $D$ and its corresponding affine dynamics. The DBM $\textbf{D}$ corresponds to variables $x_1,x_2,\ldots,$ and their primed version $x_1^\prime,x_2^\prime,\ldots,$. Then, the canonical-form DBM $\text{cf}(\textbf{D})$ is computed. The image of $D$ is established by removing all inequalities with non-primed variables in $\text{cf}(\textbf{D})$. This procedure has complexity $O(n^3)$ \cite{Dieky3}.
\begin{example}
\label{ex4}
Let us compute the image of $D=\{\text{\textbf{x}} \in \mathbb{R}^3|x_1-x_2\geq 6,x_1-x_3> -1,x_2-x_3\geq 2\}$ w.r.t. its affine dynamics $x_1^\prime = x_2+1,x_2^\prime = x_1+5,x_3^\prime = x_1+2$. The DBM generated from $D$ and the affine dynamics is
$\textbf{D}=\{[\text{\textbf{x}}^\top~(\text{\textbf{x}}^\prime)^\top]^\top\in \mathbb{R}^6|x_1-x_2\geq 6,x_1-x_3> -1,x_2-x_3\geq 2,x_1^\prime - x_2=1,x_2^\prime -x_1= 5,x_3^\prime -x_1=2\}$. The canonical-form representation of $\textbf D$ is $\text{cf}(\textbf{D})=\{[\text{\textbf{x}}^\top~(\text{\textbf{x}}^\prime)^\top]^\top\in \mathbb{R}^6|x_1-x_2\geq 6,x_1-x_3\geq 8 ,x_2-x_3\geq 2,x_1^\prime-x_1\leq -5, x_1^\prime - x_2=1,x_1^\prime -x_3\geq 3,x_2^\prime -x_1= 5,x_2^\prime-x_2\geq 11, x_2^\prime-x_3\geq 13,x_3^\prime -x_1=2,x_1^\prime-x_2^\prime\leq -10,x_1^\prime-x_3^\prime\leq -7,x_2^\prime-x_3^\prime=3\}$. The image of $D$ over the given affine dynamics is generated by removing all inequalities containing $x_1,x_2$ or $x_3$, i.e. $\{\text{\textbf{x}}^\prime\in \mathbb{R}^3|x_1^\prime-x_2^\prime\leq -10,x_1^\prime-x_3^\prime\leq -7,x_2^\prime-x_3^\prime=3\}.$ \QEDB
\end{example}
The above procedure can be improved by manipulating DBM $D$ directly from the affine dynamics. By \eqref{eq8}, one could write $x_i^\prime=x_{g_i}+A_g(i,g_i)$ where $x_i$ and $x_i^\prime$ represent the current and next variables, respectively. For each pair $(i,j)$, we have $x_i^\prime-x_j^\prime=x_{g_i}-x_{g_j}+A_g(i,g_i)-A_g(j,g_j)$. This relation ensures that the bound of $x_i^\prime-x_j^\prime$ can be determined uniquely from $x_{g_i}-x_{g_j}$ and $A_g(i,g_i)-A_g(j,g_j)$. 
\begin{proposition}
The image of a DBM $D$ w.r.t. affine dynamics $x_i^\prime=x_{g_i}+A_g(i,g_i)$ for $1\leq i\leq n$ is a set $D^\prime=\bigcap_{i=1}^n\bigcap_{j=1}^n\{\textbf{x}^\prime\in \mathbb{R}^{n}| x_i^\prime-x_j^\prime=x_{g_i}-x_{g_j}+A_g(i,g_i)-A_g(j,g_j)\}$, where the bound of $x_{g_i}-x_{g_j}$ is taken from $D$. \QEDB
\label{prop5}
\end{proposition}
\begin{example}
\label{ex5}
We compute the image of $D=\{\text{\textbf{x}}\in \mathbb{R}^3|x_1-x_2\geq 6,x_1-x_3> -1,x_2-x_3\geq 2\}$ with the same affine dynamics $x_1^\prime = x_2+1,x_2^\prime = x_1+5,x_3^\prime = x_1+2$. From the affine dynamics and $D$, we have 
$x_1^\prime-x_2^\prime=x_2-x_1-4\leq -10,~~x_1^\prime-x_3^\prime=x_2-x_1-1\leq -7,$ and $
x_2^\prime-x_3^\prime=3$
which yields a set $\{\text{\textbf{x}}^\prime\in \mathbb{R}^3|x_1^\prime-x_2^\prime\leq -10,x_1^\prime-x_3^\prime\leq -7,x_2^\prime-x_3^\prime=3\}.$
\QEDB
\end{example}
Algorithm 3 shows a procedure to generate the image of $(D,S)$ w.r.t. the affine dynamics represented by $\textbf{x}^\prime=A_g\otimes \textbf{x}$. It requires DBM $(D,S)$ located in a PWA region $R_g$. This means that there is exactly one finite coefficient $g$ such that $(D,S)\subseteq R_g$. The complexity of Algorithm 3 is in $O(n^2)$ as the addition step at 4 line has complexity of $O(1)$. \vspace*{-3ex}
\begin{algorithm}[!ht]
    \SetKwInOut{Input}{Input}
    \SetKwInOut{Output}{Output}
    \Input{$(D,S)$, a DBM in $\mathbb{R}^n$\\
    $g$, the corresponding finite coefficient such that $(D,S)\subseteq R_g$\\
    $A_g$, a region matrix which represents the affine dynamics
    }
    \Output{$(D^\prime,S^\prime)$, image of $D$ w.r.t. $\textbf{x}^\prime=A_g\otimes \textbf{x}$\\
    }
    \textbf{Initialize} $(D^\prime,S^\prime)$ with $\mathbb{R}^n$\\
      \For{$i\in\{0,\ldots,n\}$ }{
            \For{$j\in\{0,\ldots,n\}$ }{
            $D^\prime(i,j):=D(g_i,g_j)+A_g(i,g_i)-A_g(j,g_j)$\\
            $S^\prime(i,j):=S(g_i,g_j)$            
      }		
      }      
    \caption{Computing the image of DBM $D$ w.r.t. $\textbf{x}^\prime=A_g\otimes \textbf{x}$}
\end{algorithm}\vspace*{-5ex}\\
As an alternative, we also show that the image of a DBM can be computed by tropical matrix multiplications with the corresponding region matrix $A_g$. 
\begin{proposition}
The image of DBM $D$ in $\mathbb{R}^n$ w.r.t. the affine dynamics $\textbf{x}^\prime=A_g\otimes \textbf{x}$ is $D^\prime=A_g\otimes D\otimes A_g^\mathsf{c}$.\QEDB
\end{proposition}
The procedure to compute the image of DBM $D$ w.r.t. MPL system can be viewed as the extension of Algorithm 3. Before applying Algorithm 3, the DBM $D$ is intersected with each region of the PWA system. Then, for each nonempty intersection we apply Algorithm 3. The worst-case complexity is $O(|\hat{R}|n^2)$ where $|\hat{R}|$ denotes the number of PWA regions.

In \cite{Dieky3}, the procedure to compute the inverse image of $D^\prime$ w.r.t. affine dynamics involves: 1) constructing DBM $\textbf{D}$ that consists of $D^\prime$ and its corresponding affine dynamics, 2) generating the canonical form of $\textbf{D}$ and 3) removing all inequalities with primed variables. The complexity of computing the inverse image using this procedure is $O(n^3)$ as it involves the emptiness checking of a DBM \cite{Dieky3}. \vspace*{-2ex}
\begin{example}
\label{ex6}
Let us compute the inverse image of $D^\prime=\{\text{\textbf{x}}^\prime\in \mathbb{R}^3|x_1^\prime-x_2^\prime\leq -10,x_1^\prime-x_3^\prime\leq -7,x_2^\prime-x_3^\prime=3\}$ w.r.t. affine dynamics $x_1^\prime = x_2+1,x_2^\prime = x_1+5,x_3^\prime = x_1+2$. The DBM generated from $D^\prime$ and the affine dynamic is
$\textbf{D}^\prime=\{[\text{\textbf{x}}^\top~(\text{\textbf{x}}^\prime)^\top]^\top\in \mathbb{R}^6|x_1^\prime-x_2^\prime\leq -10,x_1^\prime-x_3^\prime\leq -7,x_2^\prime-x_3^\prime=3,x_1^\prime - x_2=1,x_2^\prime -x_1= 5,x_3^\prime -x_1=2\}$.
The canonical-form of $\textbf{D}$ is 
$\text{cf}(\textbf{D)}=\{[\text{\textbf{x}}^\top~(\text{\textbf{x}}^\prime)^\top]^\top\in \mathbb{R}^6|x_1-x_2\geq 6, x_1^\prime-x_1\leq -5, x_1^\prime - x_2=1,x_1^\prime -x_3\geq 3,x_2^\prime -x_1= 5,x_2^\prime-x_2\geq 11, x_3^\prime -x_1=2,x_3^\prime-x_2\geq 8,x_1^\prime-x_2^\prime\leq -10,x_1^\prime-x_3^\prime\leq -7,x_2^\prime-x_3^\prime=3\}$. The inverse image of $D^\prime$ over the given affine dynamic is computed by removing all inequalities containing $x_1^\prime,x_2^\prime$ or $x_3^\prime$, i.e. $\{\text{\textbf{x}}\in \mathbb{R}^3|x_1-x_2\geq 6 \}.$ \QEDB
\end{example}
The inverse image of $D^\prime$ can be established by manipulating $D^\prime$ from the affine dynamics. Notice that, from \eqref{eq8}, we have $x_{g_i}-x_{g_j}=x_i^\prime-x_j^\prime+A_g(j,g_j)-A_g(i,g_i)$. Unlike the previous case, it is possible that $x_{g_i}-x_{g_j}$ has multiple bounds. This happens because there is a case $g_{i_1}=g_{i_2}$ but $i_1\neq i_2$. In this case, the bound of $x_{g_i}-x_{g_j}$ is taken from the tightest bound among all possibilities.
\begin{proposition}
The inverse image of DBM $D^\prime$ w.r.t. affine dynamics $x_i^\prime=x_{g_i}+A_g(i,g_i)$ for $i\in\{1,\ldots ,n\}$ is a set $D=\bigcap_{i=1}^n\bigcap_{j=1}^n\{\textbf{x}^\prime\in \mathbb{R}^{n}| x_{g_i}-x_{g_j}=x_i^\prime-x_j^\prime+A_g(j,g_j)-A_g(i,g_i)\}$ where the bound of $x_{i}^\prime-x_{j}^\prime$ is taken from $D^\prime$.  \QEDB
\label{prop6}
\end{proposition}
Algorithm 4 shows the steps to compute the inverse image of DBM $D^\prime$ over the affine dynamics $\textbf{x}^\prime=A_g\otimes \textbf{x}$. It has similarity with Algorithm 3 except it updates the value of $D(g_i,g_j)$ and $S(g_i,g_j)$ for all $i,j\in\{0,\ldots,n\}$. The variables $b$ and $s$ in lines 4-5 represent the new bound of $x_{g_i}-x_{g_j}$; that is, $x_{g_i}-x_{g_j}\geq b$ if $s=1$ and $x_{g_i}-x_{g_j}> b$ if $s=0$. If the new bound is larger then it replaces the old one. In case of they are equal, we only need to update the operator. 
\vspace*{-3ex}
\begin{algorithm}[!ht]
    \SetKwInOut{Input}{Input}
    \SetKwInOut{Output}{Output}
    \Input{$(D^\prime,S^\prime)$, a DBM in $\mathbb{R}^n$\\
    $g$, the corresponding finite coefficient such that $(D,S)\subseteq R_g$\\
    $A_g$, a region matrix which represents the affine dynamics
    }
    \Output{$(D,S)$, inverse image of $D$ w.r.t. $\textbf{x}^\prime=A_g\otimes \textbf{x}$\\
    }
    \textbf{Initialize} $(D,S)$ with $\mathbb{R}^n$\\
      \For{$i\in\{0,\ldots,n\}$ }{
            \For{$j\in\{0,\ldots,n\}$ }{
            $b:=D^\prime(i,j)+A_g(j,g_j)-A_g(i,g_i)$\\
            $s:=S^\prime(i,j)$\\
            \uIf{$b>D(g_i,g_j)$}
            {$D(g_i,g_j):=b$\\
            $S(g_i,g_j):=s$\\}
            \uElseIf{$b=D(g_i,g_j)$}
            {$S(g_i,g_j):=\min\{s,S(g_i,g_j)\}$}
		\textbf{end}
      }		
      }         
    \caption{
    Computing the inverse image of DBM $D^\prime$ w.r.t. $\textbf{x}^\prime=A_g\otimes \textbf{x}$}
\end{algorithm}\vspace*{-3ex}\\
Similar to Algorithm 3, Algorithm 4 has complexity in $O(n^2)$. In tropical algebra, the procedure of Algorithm 4 can be expressed as tropical matrix multiplications using a region matrix and its conjugate.
\begin{proposition}
The inverse image of DBM $D^\prime$ in $\mathbb{R}^n$ w.r.t. affine dynamic $\textbf{x}^\prime=A_g\otimes \textbf{x}$ is $D =(A_g^\mathsf{c}\otimes D^\prime\otimes A_g)\oplus I_{n+1}$.\QEDB
\end{proposition}
The procedure to compute the inverse image of DBM $D^\prime$ w.r.t. MPL system can be viewed as the extension of Algorithm 4.
First, we compute the inverse image of DBM $D^\prime$ w.r.t.\ all affine dynamics. Then each inverse image is intersected with the corresponding PWA region. The worst-case complexity is $O(|\hat{R}|n^2)$. 
\subsection{Generating the Abstract Transitions}
As we mentioned before, the transition relations are generated by one-step forward-reachability analysis, and involve the image computation of each abstract state. Suppose $\hat{R}=\{\hat{r}_1,\ldots,\hat{r}_{|\hat{R}|}\}$\footnote{$\hat{R}$ is the collection of non-empty $R_g$. We use small letter $\hat{r}_i$ for sake of simplicity.} is the set of abstract states generated by Algorithm 2. There is a transition from $\hat{r}_i$ to $\hat{r}_j$ if $\mathsf{Im}(\hat{r}_i)\cap \hat{r}_j\neq \emptyset$, where $\mathsf{Im}(\hat{r}_i)=\{A\otimes \textbf{x}|\textbf{x}\in \hat{r}_i\}$ which can be computed by Algorithm 3. Notice that, each abstract state corresponds to an unique affine dynamics. The procedure to generate the transitions is summarized in Algorithm 5.
\vspace*{-4ex}
\begin{algorithm}[!ht]
    \SetKwInOut{Input}{Input}
    \SetKwInOut{Output}{Output}
    \Input{$\hat{R}=\{\hat{r}_1,\ldots,\hat{r}_{|\hat{R}|}\}$, the set of abstract states generated by Algorithm 2
    }
    \Output{$T\subseteq \hat{R}\times\hat{R}$, a transition relation
    }
    \textbf{Initialize} $T$ with an empty set\\
      \For{$i\in\{1,\ldots,|\hat{R}|\}$ }{
            \For{$j\in\{1,\ldots,|\hat{R}|\}$ }{
            compute $\mathsf{Im}(\hat{r}_i)$ by Algorithm 3\\
            \If{$\mathsf{Im}(\hat{r}_i)\cap \hat{r}_j\neq \emptyset$}
            {$T:=T\cup \{(\hat{r}_i,\hat{r}_j)\}$}
      }
      }

    \caption{
    Generating the transition via one-step forward-reachability analysis }
\end{algorithm}\vspace*{-4ex}\\
Algorithm 5 spends most time for emptiness checking at line 5. Therefore, the worst-case complexity is in $O(n^3|\hat{R}|^2)$, where $n$ is the dimension of tropical matrix $A$ in Algorithm 2.

\begin{figure}[t]
\centering
\begin{tikzpicture}[node distance=0.7cm and 0.7cm,scale=0.4]
\fill[black!35!white] (3,2) -- (7,2)--(7,5)--(6,5)--(3,2);
\fill[black!40!white] (3,2) -- (-1,2)--(-1,5)--(6,5)--(3,2);
\fill[black!35!white] (-1,2)--(-1,5)--(-5,5)--(-5,2)--(-1,2);
\fill[black!40!white] (3,2) -- (7,2)--(7,-7)--(-1,-7)--(-1,-2)--(3,2);
\fill[black!35!white] (-1,2)--(3,2)--(-1,-2)--(-1,2);
\fill[black!35!white] (-1,-2)--(-1,-7)--(-5,-7)--(-5,-6);
\fill[black!40!white] (-1,2)--(-5,2)--(-5,-6)--(-1,-2);
\draw[->,>=stealth] (-5,0) -- (7,0);
\draw[->,>=stealth] (0,-7) -- (0,5);
\draw[-] (-5,-6) -- (6,5);
\draw[-,dashed] (-4.85-0.12,-5.85) -- (6-0.12,5);
\draw[-] (-5,2) -- (7,2);
\draw[-,dashed] (-5,2-0.1) -- (7,2-0.1);
\draw[-] (-1,-7) -- (-1,5);
\draw[-,dashed] (-1+0.1,-7) -- (-1+0.1,5);
\foreach \x in {-4,...,6}{
	\draw[-] (\x,0.1)--(\x,-0.1);
	\draw[-] (0.1,-\x)--(-0.1,-\x);
	}
	\node at (3,-0.4) {\scriptsize 3};
	\node at (-0.3,3) {\scriptsize 3};
	
	\node at (-3,3.5) {$\boldsymbol{\hat{r}_3}$};
	\node at (3.5,-3.5) {$\boldsymbol{\hat{r}_4}$};
	\node at (0.8,0.8) {$\boldsymbol{\hat{r}_5}$};
	\node at (-2.7,-5.3) {$\boldsymbol{\hat{r}_6}$};
	\node at (-3,-1) {$\boldsymbol{\hat{r}_7}$};
	\node at (7.55,0) {$x_1$};
	\node at (0,5.3) {$x_2$};
	
	\node at (5.8,3.3) {$\boldsymbol{\hat{r}_1}$};
	\node at (2,3.5) {$\boldsymbol{\hat{r}_2}$};
	\node at (1,-8) {$(a)$};
\end{tikzpicture} ~~~~~~~
 \begin{tikzpicture}[node distance=0.7cm and 0.7cm]\footnotesize
\tikzstyle{place}=[circle,thick,draw=black,minimum size=2mm,scale=0.65]
   \node[place] (a)   {$\hat{r}_4$};
   \node[place] (b) [left=of a,xshift=-1cm] {$\hat{r}_1$};
   \node[place] (c) [below =of a,yshift=-0.8cm ] {$\hat{r}_2$};
   \node[place] (d) [left =of c,xshift=-1cm] {$\hat{r}_7$};
   \node[place] (e) [below=of c,yshift=-0.8cm] {$\hat{r}_6$};
   \node[place] (f) [below=of d,yshift=-0.8cm ] {$\hat{r}_5$};
   \node[place] (g) [below=of f,yshift=1cm,xshift=-1cm ] {$\hat{r}_3$};

\draw[->,>=stealth] (e) to (f);
\draw[->,>=stealth] (e) to (d);
\draw[->,>=stealth] (e) [bend right=20] to (c);
\draw[->,>=stealth] (c) [bend right=20] to (e);
\draw[->,>=stealth] (f) [bend right=20] to (d);
\draw[->,>=stealth] (d) [bend right=20] to (f);
\draw[->,>=stealth] (d) [bend right=20] to (c);
\draw[->,>=stealth] (c) [bend right=20] to (d);
\draw[->,>=stealth] (g) [bend left=20] to (d);
\draw[->,>=stealth] (g) [bend right=20] to (e);
\draw[->,>=stealth] (d)  [loop left] to (d);
\draw[->,>=stealth] (a) to (d);
\draw[->,>=stealth] (b) to (d);
\node [below=of g,yshift=0.5cm,xshift=1.5cm ] {$(b)$};
\end{tikzpicture} 
\caption{$(a)$ Plot of partitions (and corresponding abstract states), projected on the plane $x_3=0$. 
The solid and dashed lines represent $\geq$ and $>$, respectively. $(b)$ Transition relations among abstract states.}
\label{figu4}
\end{figure}
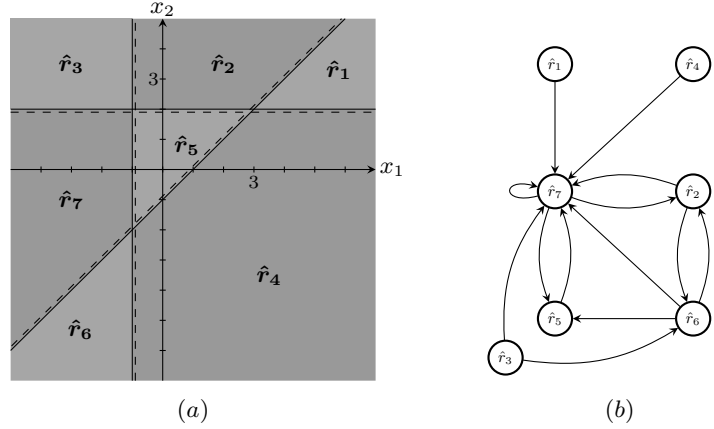

\begin{example}
The tropical matrix in Example 1 has $2^3=8$ finite coefficients. The resulting abstract states generated by Algorithm 2 are
$\hat{r}_{1}=\{\textbf{x}\in\mathbb{R}^3|x_1-x_2\geq 1,x_1-x_3\geq 3,x_2-x_3\geq 2\},
\hat{r}_{2}=\{\textbf{x}\in\mathbb{R}^3| x_1-x_2< 1,x_1-x_3> -1,x_2-x_3\geq 2\},
\hat{r}_{3}=\{\textbf{x}\in\mathbb{R}^3| x_1-x_2\leq -3,x_1-x_3\leq -1,x_2-x_3\geq 2\},
\hat{r}_{4}=\{\textbf{x}\in\mathbb{R}^3| x_1-x_2\geq 1, x_1-x_3> -1, x_2-x_3< 2\},
\hat{r}_{5}=\{\textbf{x}\in\mathbb{R}^3| -3< x_1-x_2< 1,-1< x_1-x_3< 3, -2< x_2-x_3< 2\},
\hat{r}_{6}=\{\textbf{x}\in\mathbb{R}^3| x_1-x_2\geq 1, x_1-x_3\leq -1, x_2-x_3\leq -2\},$ and $
\hat{r}_{7}=\{\textbf{x}\in\mathbb{R}^3| x_1-x_2< 1,x_1-x_3\leq -1, x_2-x_3< 2\},$
which correspond to finite coefficients $(2,1,1), (2,1,2), (2,3,2),(3,1,1),(3,1,2),$ $(3,3,1),$ and $(3,3,2)$, respectively. The only finite coefficient that leads to an empty set is $(2,3,1)$. \autoref{figu4} shows the illustrations of abstract states and transition relations.

\end{example}
\section{Computational Benchmarks}
We compare the run-time of abstraction algorithms in this paper with the procedures in \textsf{VeriSiMPL} 1.4 \cite{Dieky4}. For increasing $n$, we generate matrices $A\in\mathbb{R}_{\max}^{n\times n}$ with two finite elements in each row, with value ranging between 1 and 100. The location and value of the finite elements are chosen randomly. The computational benchmark has been implemented on a high-performance computing cluster at the University of Oxford \cite{Andrew}. 
\newpage
\vspace*{-6ex}
We run the experiments for both procedures (\textsf{VeriSiMPL} 1.4 and Tropical) using MATLAB R2017a with parallel computing. Over 10 different MPL systems for each dimension, \autoref{tab1} shows the running time to generate the abstract states and transitions. Each entry represents the average and maximal values.
\vspace*{-6ex}
\begin{table}[!ht]
\centering
\captionsetup{format=hang,width=.8\linewidth}
\caption{Generation of abstract states and transitions }
\label{tab1}
\begin{scriptsize}
\begin{tabular}{|c|r|r|r|r|}
\hline
   & \multicolumn{2}{c|}{\textsf{VeriSiMPL} 1.4.}&\multicolumn{2}{c|}{Tropical}\\
   \cline{2-5}
       &\multicolumn{1}{c|}{time for}&\multicolumn{1}{c|}{time for} &\multicolumn{1}{c|}{time for} &\multicolumn{1}{c|}{time for} \\
  $n$  &\multicolumn{1}{c|}{generating} &\multicolumn{1}{c|}{generating} &\multicolumn{1}{c|}{generating}&\multicolumn{1}{c|}{generating}\\
        &\multicolumn{1}{c|}{abstract states}&\multicolumn{1}{c|}{transitions}&\multicolumn{1}{c|}{abstract states}&\multicolumn{1}{c|}{transitions}\\
    \hline
    3&$\{  7.51,	9.82\}[\text{ms}]$&$\{0.13,	0.21\}[\text{sec}]$&$\{4.04,	8.39\}[\text{ms}]$&$\{		0.12,	0.17\}[\text{sec}]$\\\hline
4&$\{  11.29,	15.58\}[\text{ms}]$&$\{	0.20,	0.29\}[\text{sec}]$&$\{	5.23,	16.10\}[\text{ms}]$&$\{		0.17,	0.22\}[\text{sec}]$\\\hline
5&$\{  18.51,	28.19\}[\text{ms}]$&$\{	0.20,	0.21\}[\text{sec}]$&$\{		5.16,	6.89	\}[\text{ms}]$&$\{	0.19,	0.20\}[\text{sec}]$\\\hline
6&$\{  49.22,	55.10\}[\text{ms}]$&$\{	0.21,	0.22\}[\text{sec}]$&$\{		9.99,	11.44\}[\text{ms}]$&$\{		0.20,	0.21\}[\text{sec}]$\\\hline
7&$\{  90.88,	118.94\}[\text{ms}]$&$\{	0.24,	0.26\}[\text{sec}]$&$\{		15.88,	20.67\}[\text{ms}]$&$\{		0.22,	0.24\}[\text{sec}]$\\\hline
 8&$\{  0.21,	0.28\}[\text{sec}]$&$\{0.32,	0.44\}[\text{sec}]$&$\{		0.04	,0.04\}[\text{sec}]$&$\{		0.27,	0.38\}[\text{sec}]$\\\hline
9&$\{  0.52,	0.69\}[\text{sec}]$&$\{0.72,	1.07\}[\text{sec}]$&$\{		0.07,	0.10\}[\text{sec}]$&$\{		0.60,	0.91\}[\text{sec}]$\\\hline
10&$\{  1.25,	1.88\}[\text{sec}]$&$\{2.62,	4.48\}[\text{sec}]$&$\{		0.14,	0.17\}[\text{sec}]$&$\{		2.38	,4.22\}[\text{sec}]$\\\hline
11&$\{  3.87,	5.14\}[\text{sec}]$&$\{17.62,	29.44\}[\text{sec}]$&$\{		0.35,	0.39\}[\text{sec}]$&$\{		17.17,	28.88\}[\text{sec}]$\\\hline
12&$\{  8.34,	14.22\}[\text{sec}]$&$\{1.20,	2.24\}[\text{min}]$&$\{		0.61,	0.71\}[\text{sec}]$&$\{		1.10,	2.19\}[\text{min}]$\\\hline
13&$\{  26.17,	45.17\}[\text{sec}]$&$\{	5.05,	10.45\}[\text{min}]$&$\{		1.21,	1.37\}[\text{sec}]$&$\{		4.98,	10.40\}[\text{min}]$\\\hline
14&$\{1.81,	4.24\}[\text{min}]$&$\{	41.14,	112.09\}[\text{min}]$&$\{	0.06,	0.07\}[\text{min}]$&$\{	40.61,	110.06\}[\text{min}]$\\\hline
15&$\{10.29,	23.18\}[\text{min}]$&$\{	2.63,	7.57\}[\text{hr}]$&$\{	0.11,	0.17\}[\text{min}]$&$\{	2.57,	7.65\}[\text{hr}]$\\\hline
\end{tabular}
\end{scriptsize}
\end{table}\vspace*{-4ex}

With regards to the generation of abstract states, the tropical algebra based algorithm is much faster than \textsf{VeriSiMPL} 1.4. As the dimension increases, we see an increasing gap of the running time. For a 12-dimensional MPL system over 10 independent experiments, the time needed to compute abstract states using tropical based algorithm is less than 1 second. In comparison, average running time using \textsf{VeriSiMPL} 1.4 for the same dimension is 8.34 seconds. 

For the generation of transitions, the running time of tropical algebra-based algorithm is slightly faster than that of \textsf{VeriSiMPL} 1.4. We remind that the procedure to generate transitions involves the image computation of each abstract state. In comparison to the second and fourth columns of \autoref{tab1}, \autoref{tab2} shows the running time to compute the image of abstract states. Each entries represents the average and maximum of running time. It shows that our proposed algorithm for image computation of DBMs is faster than \textsf{VeriSiMPL} 1.4.\vspace*{-6ex}
\begin{table}[!ht]
\centering
\begin{scriptsize}
\captionsetup{format=hang,width=.7\linewidth}
\caption{Computation of the image of abstract states }
\label{tab2}
\begin{tabular}{|c|r|r|}
\hline
$n$& \multicolumn{1}{c|}{\textsf{VeriSiMPL} 1.4.}&\multicolumn{1}{c|}{Tropical}\\\hline
3&$\{ 0.84,	1.13\}[\text{ms}]$&$\{	0.16,	0.23\}[\text{ms}]$\\\hline
4&$\{ 1.13,	1.76\}[\text{ms}]$&$\{	0.13,	0.20\}[\text{ms}]$\\\hline
5&$\{ 1.53,	2.40\}[\text{ms}]$&$\{	0.14,	0.16\}[\text{ms}]$\\\hline
6&$\{ 5.32,	6.68\}[\text{ms}]$&$\{	0.18,	0.20\}[\text{ms}]$\\\hline
7&$\{ 11.22,	15.19\}[\text{ms}]$&$\{	0.31,	0.44\}[\text{ms}]$\\\hline
8&$\{ 26.05,	46.94\}[\text{ms}]$&$\{	0.71,	1.19\}[\text{ms}]$\\\hline
9&$\{ 70.31,	92.87\}[\text{ms}]$&$\{	2.37,	3.37\}[\text{ms}]$\\\hline
10&$\{ 153.07,	183.08\}[\text{ms}]$&$\{	4.06,	6.57\}[\text{ms}]$\\\hline
11&$\{ 380.01,	477.94\}[\text{ms}]$&$\{	5.58,	8.19\}[\text{ms}]$\\\hline
12 &$\{0.79	,1.13\}[\text{sec}]$&$\{	0.02,	0.03\}[\text{sec}]$\\\hline
13&$\{ 1.96,	3.13\}[\text{sec}]$&$\{	0.03,	0.04\}[\text{sec}]$\\\hline
14&$\{ 5.51,	9.60\}[\text{sec}]$&$\{		0.06,	0.16\}[\text{sec}]$\\\hline
15&$\{ 14.33,	23.82\}[\text{sec}]$&$\{		0.49,	0.87\}[\text{sec}]$\\\hline
\end{tabular}
\end{scriptsize}
\end{table}\vspace*{-4ex}\\
\indent We also compare the running time algorithms when applying forward- and backward-reachability analysis. We generate the forward reach set \cite[Def 4.1]{Dieky1} and backward reach set \cite[Def 4.3]{Dieky1} from an initial and a final set, respectively. 
In more detail, suppose $\mathcal{X}_0$ is the set of initial conditions; 
the forward reach set $\mathcal{X}_k$ is defined recursively as the image of $\mathcal{X}_{k-1}$, namely 
$$
\mathcal{X}_k =\{A\otimes \textbf{x}| \textbf{x}\in \mathcal{X}_{k-1}\}.
$$
On the other hand, suppose $\mathcal{Y}_0$ is a set of final conditions. The backward reach set $\mathcal{Y}_{-k}$ is defined via the inverse image of $\mathcal{Y}_{-k+1},$
$$\mathcal{Y}_{-k} =\{\textbf{y}\in \mathbb{R}^n| A\otimes \textbf{y}\in \mathcal{Y}_{-k+1}\},$$
where $n$ is the dimension of $A$. 

We select $\mathcal{X}_0 =\{\textbf{x}\in\mathbb{R}^n:0\leq x_1\leq 1,\ldots,0\leq x_n\leq 1\}$ and $\mathcal{Y}_0 =\{\textbf{y}\in\mathbb{R}^n:90\leq y_1\leq 100,\ldots,90\leq y_n\leq 100\}$ as the sets of initial and final conditions, respectively. 
The experiments have been implemented to compute the forward reach sets $\mathcal{X}_1,\ldots,\mathcal{X}_N$ and the backward reach sets $\mathcal{Y}_{-1},\ldots,\mathcal{Y}_{-N}$ for $N=10$. 
Notice that it is possible that the inverse image of $\mathcal{Y}_{-k+1}$ results in an empty set: 
in this case, the computation of backward reach sets is terminated, since $\mathcal{Y}_{-k }=\ldots=\mathcal{Y}_{-N}=\emptyset$. 
(If this termination happens, it applies for both \textsf{VeriSiMPL} 1.4 and the algorithms based on tropical algebra.) 

\autoref{tab3} reports the average computation of PWA system and reach sets over 10 independent experiments for each dimension. 
In general, algorithms based on tropical algebra outperform those of \textsf{VeriSiMPL} 1.4. 
For a 15-dimensional MPL system, the average time to generate PWA system using \textsf{VeriSiMPL} 1.4 is just over 20 seconds. In comparison, the computation time for tropical algorithm is under 5 seconds.
\begin{table}[!ht]
\centering
\caption{Reachability analysis }
\label{tab3}
\begin{scriptsize}
\begin{tabular}{|c|r|r|r|r|r|r|}
\hline
   & \multicolumn{3}{c|}{\textsf{VeriSiMPL} 1.4.}&\multicolumn{3}{c|}{Tropical}\\
   \cline{2-7}
       &\multicolumn{1}{c|}{time for}&\multicolumn{1}{c|}{time for} &\multicolumn{1}{c|}{time for} &\multicolumn{1}{c|}{time for} &\multicolumn{1}{c|}{time for} &\multicolumn{1}{c|}{time for} \\
$n$ &\multicolumn{1}{c|}{generating} &\multicolumn{1}{c|}{generating} &\multicolumn{1}{c|}{generating}&\multicolumn{1}{c|}{generating}&\multicolumn{1}{c|}{generating}&\multicolumn{1}{c|}{generating}\\
        &\multicolumn{1}{c|}{PWA}&\multicolumn{1}{c|}{ forward }&\multicolumn{1}{c|}{backward }&\multicolumn{1}{c|}{PWA }&\multicolumn{1}{c|}{ forward}&\multicolumn{1}{c|}{backward }\\
        &\multicolumn{1}{c|}{system}&\multicolumn{1}{c|}{reach sets}&\multicolumn{1}{c|}{reach sets}&\multicolumn{1}{c|}{system}&\multicolumn{1}{c|}{reach sets}&\multicolumn{1}{c|}{ sets}\\
    \hline
3&$ 2.55[\text{ms}]	$&$11.37[\text{ms}]$&$	5.73[\text{ms}]$&$	1.70	[\text{ms}]$&$8.33[\text{ms}]$&$	5.63[\text{ms}]$\\\hline
4&$ 4.31[\text{ms}]	$&$9.87[\text{ms}]$&$	27.00[\text{ms}]$&$	1.37	[\text{ms}]$&$7.72[\text{ms}]$&$	28.48[\text{ms}]$\\\hline
5 &$9.23[\text{ms}]	$&$11.77[\text{ms}]$&$	3.62[\text{ms}]$&$	1.88[\text{ms}]	$&$9.25[\text{ms}]$&$	2.89[\text{ms}]$\\\hline
6 &$23.44[\text{ms}]	$&$18.49[\text{ms}]$&$	9.76[\text{ms}]	$&$3.80[\text{ms}]$&$	13.81[\text{ms}]	$&$7.35[\text{ms}]$\\\hline
7 &$49.59[\text{ms}]	$&$35.68[\text{ms}]$&$	21.53[\text{ms}]$&$	7.84[\text{ms}]$&$	32.02[\text{ms}]	$&$17.92[\text{ms}]$\\\hline
8 &$108.75 [\text{ms}]$&$	85.27[\text{ms}]	$&$34.05[\text{ms}]$&$	16.84[\text{ms}]	$&$73.63[\text{ms}]$&$	28.62[\text{ms}]$\\\hline
9 &$0.25[\text{sec}]	$&$0.18[\text{sec}]$&$	0.09[\text{sec}]$&$	0.03[\text{sec}]	$&$0.17[\text{sec}]$&$	0.07[\text{sec}]$\\\hline
10&$ 0.48[\text{sec}]$&$	0.28[\text{sec}]$&$ 	0.17[\text{sec}]	$&$0.08[\text{sec}]$&$	0.25[\text{sec}]$&$	0.14[\text{sec}]$\\\hline
11 &$1.19[\text{sec}]	$&$0.77[\text{sec}]$&$	1.35[\text{sec}]$&$	0.18[\text{sec}]$&$	0.76[\text{sec}]	$&$1.13[\text{sec}]$\\\hline
12 &$2.52[\text{sec}]	$&$1.14[\text{sec}]$&$	0.88[\text{sec}]$&$	0.38[\text{sec}]	$&$1.01[\text{sec}]$&$	0.70[\text{sec}]$\\\hline
13 &$7.02[\text{sec}]	$&$3.96[\text{sec}]$&$	2.78[\text{sec}]$&$	1.09[\text{sec}]	$&$3.56[\text{sec}]$&$	1.95[\text{sec}]$\\\hline
14 &$8.15[\text{sec}]	$&$5.54[\text{sec}]$&$	4.61[\text{sec}]$&$	1.54[\text{sec}]	$&$5.24[\text{sec}]$&$	2.98[\text{sec}]$\\\hline
15 &$20.60[\text{sec}]$&$	19.23[\text{sec}]	$&$12.39[\text{sec}]	$&$4.21[\text{sec}]	$&$18.37[\text{sec}]	$&$7.16[\text{sec}]$\\\hline
16 &$46.92[\text{sec}]	$&$60.19[\text{sec}]$&$	36.00[\text{sec}]$&$	9.62[\text{sec}]	$&$58.70[\text{sec}]$&$	20.41[\text{sec}]$\\\hline
18 &$2.98[\text{min}]	$&$3.91[\text{min}]$&$	2.61[\text{min}]$&$	0.83[\text{min}]	$&$3.83[\text{min}]$&$	1.35[\text{min}]$\\\hline
20 &$15.74[\text{min}]	$&$21.03[\text{min}]$&$	15.21[\text{min}]$&$	4.84[\text{min}]	$&$20.86[\text{min}]$&$	7.51[\text{min}]$\\\hline
\end{tabular}
\end{scriptsize}
\end{table}

\indent Tropical algorithms also show advantages to compute reach sets. As shown in \autoref{tab3}, the average computation time for forward and backward-reachability analysis is slightly faster when using tropical procedures. There is evidence that the average time to compute the backward reach sets decreases as the dimension increases. This happens because the computation is terminated earlier once there is a $k\leq N$ such that $\mathcal{Y}_{-k}=\emptyset.$ Notice that, this condition occurs for both \textsf{VeriSiMPL} 1.4 and the new algorithms based on tropical algebra.
\section{Conclusions} 

This paper has introduced the concept of MPL abstractions using tropical operations. 
We have shown that the generation of abstract states is related to the row-definite form of the given matrix. 
The computation of image and inverse image of DBMs over the affine dynamics has also been improved based on tropical algebra operations.

The procedure has been implemented on a numerical benchmark and compared with \textsf{VeriSiMPL} 1.4. 
Algorithm 2 has showed a strong advantage to generate the abstract states especially for high-dimensional MPL systems. 
Algorithms (Algorithms 3-5) for the generation of transitions and for reachability analysis also display an improvement. 

For future research, the authors are interested to extend the tropical abstractions for non-autonomous MPL systems \cite{Bacelli}, 
with dynamics that are characterised by non-square tropical matrices.
\section*{Acknowledgements} 
The authors would like to acknowledge the use of the Advanced Research Computing (ARC) facility at the University of Oxford in carrying out the computational benchmark of this work.  The first author is supported by Indonesia Endowment Fund for Education (LPDP), while the third acknowledges the support of the Alan Turing Institute, London, UK.

\newpage
\newtheorem{innercustomprop}{Proposition}
\newenvironment{customprop}[1]
  {\renewcommand\theinnercustomprop{#1}\innercustomprop}
  {\endinnercustomprop}
\newtheorem{innercustomlem}{Lemma}
\newenvironment{customlem}[1]
  {\renewcommand\theinnercustomlem{#1}\innercustomlem}
  {\endinnercustomlem}

\begin{subappendices}
\renewcommand{\thesection}{\Alph{section}}%
\section{Appendix}
\subsection{Proof of Proposition 1}
\begin{customprop}{1}
The column-definite and row-definite form of $A\in \mathbb{R}_{\max}^{n\times n}$ w.r.t. a finite coefficient $g$ is  $\overline{A}_g=A\otimes A_g^\mathsf{c}$ and $_g\overline{A}=A_g^\mathsf{c}\otimes A $, respectively.\\

\noindent \textbf{Proof.} Let us start by writing down $[A\otimes A_\alpha^\mathsf{c}](i,j) =\bigoplus_{k=1}^n A(i,k)\otimes A_\alpha^\mathsf{c}(k,j)$. Notice that, $A_\alpha^\mathsf{c}$ is the conjugate of region matrix indexed by $\alpha$. Therefore, there exists exactly one $k$ such that $A_\alpha^\mathsf{c}(k,j)\neq \varepsilon$ for a fixed $j\in\{1,\ldots,n\}$; that is, $k=\alpha(j)$. Thus, $[A\otimes A_\alpha^\mathsf{c}](i,j) =A(i,\alpha(j))\otimes A_\alpha^\mathsf{c}(\alpha(j),j)=A(i,\alpha(j))-A_\alpha(j,\alpha(j))=\overline{A}_\alpha(i,j)$. The proof for the row-definite form is similar to that of the column-definite form. \QEDB
\end{customprop}

\subsection{Proof of Proposition 2}
\begin{customprop}{2}
The intersection of DBM $D_1$ and $D_2$ is equal to $D_1\oplus D_2$.\vspace*{2ex}\\
\textbf{Proof:} Suppose $D_1(i,j)$ and $D_2(i,j)$ corresponds to the bounds $ c_{i,j}$ and $d_{i,j}$, respectively. For each pair $(i,j)$, one needs to find the tighter bound between $c_{i,j}$ and $d_{i,j}$. The tighter bound is equal to the larger one i.e. $\max\{c_{i,j},d_{i,j}\}=c_{i,j}\oplus d_{i,j}$. \QEDB
\end{customprop}
\subsection{Proof of Proposition 3}
\begin{customprop}{3}
Given a DBM $D$, the canonical form of $D$ is $\mathsf{cf}(D)=\bigoplus_{m=0}^{n+1} D^{\otimes m}$ where $n$ is the number of variables excluding $x_0$. \\

\noindent \textbf{Proof:} The element $D(i,j)$ stores the bound for $x_i-x_j$. We show the proof by using the precedence graph $\mathcal{G}(D)$. Notice that, $[D^{\otimes m}](i,j)$ is equal to the maximal total weights of a path with length $m$ from $j$ to $i$ in $\mathcal{G}(D)$. Furthermore, $[\bigoplus_{m=0}^{n+1} D^{\otimes m}](i,j)$ is equal to the maximal total weights of a path from $j$ to $i$. Thus, $[\bigoplus_{m=0}^{n+1}] D^{\otimes m} (i,j)$ is the tightest bound for $x_i-x_j$. \QEDB
\end{customprop}
\subsection{Proof of Proposition 4}
\begin{customprop}{4}
Suppose $D$ is a canonical DBM. If $D$ is not empty then it is definite.\vspace*{2ex}\\
\noindent\textbf{Proof:} Let $D$ be a DBM in $\mathbb{R}^n$. Because $D$ is a canonical non-empty DBM, we may assume that all diagonal elements of $D$ are zero. Recall that, the tightest possible bound for $x_i-x_i$ is zero for $i\in\{0,\ldots,n\}$.  We need to show that $\text{per}(D)=0$.

Notice that, the condition $\text{per}(D)>0$ relates to the existence of circuit in $\mathcal{G}(D)$ with positive weight. The existence of such circuit leads to the emptiness of $D$. Therefore, $\text{per}(D)\leq 0$.

On the other hand, as all diagonal elements of $D$ are zero, we have $\text{per}(D)\geq \bigotimes_{i=0}^n D(i,i)=0$. Hence, we can conclude that $\text{per}(D)=0$ with identity permutation is one of the maximal permutations. \QEDB
\end{customprop}
\subsection{Proof of Proposition 5}
\begin{customprop}{5}
For each finite coefficient $g$, $R_g=~_g\overline{A}\oplus I_n$.

\noindent\textbf{Proof:} Notice that, the value of $A(i,j)-A(i,g_i)$ in \eqref{eq7} corresponds to $R_g(g_i,j)$. Furthermore, we have $$_g\overline{A}(g_i,j)=\bigoplus_{i^\ast}(A(i^\ast,j)\otimes A(i^\ast,g_i)^{\otimes -1})=\bigoplus_{i^\ast}(A(i^\ast,j)-A(i^\ast,g_i)),$$ where $i^\ast\in\{1,\ldots,n\}$ such that $g_{i^\ast}=g_i$. In the inequality part of \eqref{eq7}, one may find the multiple bounds for $x_{g_i}-x_j$. This happens whenever $g$ is not a permutation. In that case, the bound for $x_{g_i}-x_j$ is the maximum value of all corresponding bounds. Thus, $R_g(g_i,j)= \bigoplus_{i^\ast}(A(i^\ast,j)-A(i^\ast,g_i))=~_g\overline{A}(g_i,j)$ for all $i,j\in\{1,\ldots,n\}$. From here we cannot write $R_g=~_g\overline{A}$ because $_g\overline{A}$ admits an infinite diagonal element while $R_g$ is not. However, all diagonal elements in $_g\overline{A}\oplus I_n$ are 0. Therefore, $R_g=~_g\overline{A}\oplus I_n$. On the other hand, if $g$ is a permutation, we have $R_g=~_g\overline{A}=~_g\overline{A}\oplus I_n$. \QEDB
\end{customprop}
\subsection{Proof of Proposition 6}
\begin{customprop}{6}
The image of a DBM $D$ w.r.t affine dynamics $x_i^\prime=x_{g_i}+A_g(i,g_i)$ for $i\in\{1,\ldots ,n\}$ is a set characterized by $D^\prime=\bigcap_{i=1}^n\bigcap_{j=1}^n\{\textbf{x}^\prime\in \mathbb{R}^{n}| x_i^\prime-x_j^\prime=x_{g_i}-x_{g_j}+A_g(i,g_i)-A_g(j,g_j)\}$, where the bound of $x_{g_i}-x_{g_j}$ is taken from $D$. \\

\noindent \textbf{Proof:} Suppose $D^\prime$ is the image of $D$ w.r.t. the given affine dynamics. The DBM $D^\prime$ can be computed by manipulating $D$ from the affine dynamics. For each pair $(i,j)$, we have $x_i^\prime-x_j^\prime=x_{g_i}-x_{g_j}+A_g(i,g_i)-A_g(j,g_j)$. From here, we can infer that the bound for $x_i^\prime-x_j^\prime$ (primed version) corresponds to the bound of $x_{g_i}-x_{g_j}$ (non-primed  version) and a scalar $A_g(i,g_i)-A_g(j,g_j)$. Therefore, $D^\prime=\bigcap_{i=1}^n\bigcap_{j=1}^n\{\textbf{x}^\prime\in \mathbb{R}^{n}| x_i^\prime-x_j^\prime=x_{g_i}-x_{g_j}+A_g(i,g_i)-A_g(j,g_j)\}$.

\end{customprop}

\subsection{Proof of Proposition 7}
\begin{customprop}{7}
The image of DBM $D\in \mathbb{R}^n$ w.r.t. affine dynamic $\textbf{x}^\prime=A_g\otimes \textbf{x}$ is $D^\prime=A_g\otimes D\otimes A_g^\mathsf{c}$.\\
\textbf{Proof:} First, by Proposition \ref{prop5}, the image of $D$ is a set $D^\prime=\bigcap_{i=1}^n\bigcap_{j=1}^n\{\textbf{x}^\prime\in \mathbb{R}^{n}| x_i^\prime-x_j^\prime=x_{g_i}-x_{g_j}+A_g(i,g_i)-A_g(j,g_j)\}$. If $D^\prime$ is expressed as a matrix then $D^\prime(i,j)=D(g_i,g_j)+A_g(i,g_i)-A_g(j,g_j)$ for $i,j\in\{0,\ldots,n\}$ under a convention $x_0=g_0=0$ and $D(0,0)=A_g(0,0)=0$.

On the other hand, $[A_g\otimes D\otimes A_g^\mathsf{c}](i,j)=\bigoplus_{k=0}^n(A_g(i,k)\otimes (\bigoplus_{l=0}^n D(k,l)\otimes A_g^\mathsf{c}(l,j)))$. Notice that, for a fixed $j$ there is an unique $l$ such that $A_g^\mathsf{c}(l,j)\neq \varepsilon$ i.e. $l=g_j$. Similarly, for a fixed $i$, $A_g(i,k)\neq \varepsilon$ iff $k=g_i$. Therefore, $[A_g\otimes D\otimes A_g^\mathsf{c}](i,j)=A_g(i,g_i)+D(g_i,g_j)+ A_g^\mathsf{c}(g_j,j)=D(g_i,g_j)+A_g(i,g_i)-A_g(j,g_j)=D^\prime(i,j)$.\QEDB
\end{customprop}

\subsection{Proof of Proposition 8}
\begin{customprop}{8}
The inverse image of DBM $D^\prime$ w.r.t. affine dynamics $x_i^\prime=x_{g_i}+A_g(i,g_i)$ for $i\in\{1,\ldots ,n\}$ is a set characterized by $D=\bigcap_{i=1}^n\bigcap_{j=1}^n\{\textbf{x}^\prime\in \mathbb{R}^{n}| x_{g_i}-x_{g_j}=x_i^\prime-x_j^\prime+A_g(j,g_j)-A_g(i,g_i)\}$, where the bound of $x_{i}^\prime-x_{j}^\prime$ is taken from $D^\prime$.  \\

\noindent \textbf{Proof:} Similar to proof of Proposition 6. \QEDB
\end{customprop}
\subsection{Proof of Proposition 9}
\begin{customprop}{9}
The inverse image of DBM $D^\prime\in \mathbb{R}^n$ w.r.t. affine dynamic $\textbf{x}^\prime=A_g\otimes \textbf{x}$ is $D =A_g^\mathsf{c}\otimes D^\prime\otimes A_g\oplus I_{n+1}$.\\
\textbf{Proof:} Let us starts from $[A_g^\mathsf{c}\otimes D\otimes A_g](g_i,g_j)=\bigoplus_{k=0}^n(A_g^\mathsf{c}(g_i,k)\otimes (\bigoplus_{l=0}^n D(k,l)\otimes A_g(l,g_j)))$. Notice that, by \eqref{eq0}, $A_g(k,g_i)\neq \varepsilon$ if $g_k=g_i$. Thus, one can write $$\displaystyle [A_g^\mathsf{c}\otimes D\otimes A_g](g_i,g_j)=\bigoplus_{\substack{i^\ast\\j^\ast}}(-A_g(i^\ast,g_i)+D(i^\ast,j^\ast)+ A_g(j^\ast,g_j)),$$
where $i^\ast,j^\ast\in\{0,\ldots,n\}$ such that $g_{i^\ast}=g_i$ and $g_{j^\ast}=g_j$. On the other hand, by \autoref{prop6}, the inverse image of $D^\prime$ is a set $D=\bigcap_{i=1}^n\bigcap_{j=1}^n\{\textbf{x}\in \mathbb{R}^{n}| x_{g_i}-x_{g_j}=x_i^\prime-x_j^\prime+A_g(j,g_j)-A_g(i,g_i)\}$ where the bound of $x_i^\prime-x_j^\prime$ corresponds to $D^\prime (i,j)$. There are two cases. 

First, $g$ is a permutation. In this case, for each $i\in\{0,\ldots,n\}$, the value of $g_i$ is unique. Hence, from the relation $x_{g_i}-x_{g_j}=x_i^\prime-x_j^\prime+A_g(j,g_j)-A_g(i,g_i)$, the bound of $x_{g_i}-x_{g_j}$ is determined uniquely from the bound of $x_i^\prime-x_j^\prime$ plus scalar $A_g(j,g_j)-A_g(i,g_i)$; that is, $D(g_i,g_j)=D^\prime(i,j)+A_g(j,g_j)-A_g(i,g_i)$. As a consequence, we have
\begin{eqnarray}
\nonumber
\displaystyle [A_g^\mathsf{c}\otimes D\otimes A_g](g_i,g_j)&=&\bigoplus_{\substack{i^\ast\\j^\ast}}(-A_g(i^\ast,g_i)+D^\prime(i^\ast,j^\ast)+ A_g(j^\ast,g_j))\\
\nonumber &=&(-A_g(i,g_i)+D^\prime(i,j)+ A_g(j,g_j))=D(g_i,g_j)
\end{eqnarray}
Because $g_0,\ldots,g_n$ are all different, we can conclude $D =A_g^\mathsf{c}\otimes D^\prime\otimes A_g=(A_g^\mathsf{c}\otimes D^\prime\otimes A_g)\oplus I_{n+1}$.

Second, $g$ is not a permutation. Let us define $G=\{g_0,\ldots,g_n\}.$ In this case, one can find $1\leq i_{1}\neq i_2\leq n$ but $g_{i_1}=g_{i_2}$. Furthermore, there are several $w\in\{1,\ldots,n\}$ such that $w\not\in G$. For all $i^\ast$ and $j^\ast$ such that $g_i^\ast=g_i$ and $g_j^\ast=g_j$, we have $x_{g_i}-x_{g_j}=x_{i^\ast}^\prime-x_{j^\ast}^\prime+A_g(j^\ast,g_j)-A_g(i^\ast,g_i)$ which shows that $x_{g_i}-x_{g_j}$ has multiple bounds up to the number of different pairs $(i^\ast,j^\ast)$. The tightest bound of $x_{g_i}-x_{g_j}$ is equal to the maximum one; that is, $$D(g_i,g_j)=\bigoplus_{\substack{i^\ast\\j^\ast}}(-A_g(i^\ast,g_i)+D^\prime(i^\ast,j^\ast)+ A_g(j^\ast,g_j))=[A_g^\mathsf{c}\otimes D^\prime\otimes A_g](g_i,g_j).$$

\noindent From here, we have $D(i,j)=[A_g^\mathsf{c}\otimes D^\prime\otimes A_g](i,j)$ if both $i$ and $j$ are in $G$. If $i\not\in G$ or $j\not \in G$ then $D(i,j) = \varepsilon = [A_g^\mathsf{c}\otimes D^\prime\otimes A_g](i,j)$. However, as the diagonal elements of $D$ are not allowed be to non-negative, we have $D =(A_g^\mathsf{c}\otimes D^\prime\otimes A_g)\oplus I_{n+1}$.
\QEDB
\end{customprop}

\end{subappendices}

\end{document}